\documentclass[12pt]{article}

\usepackage{a4wide}
\usepackage{cite}

\usepackage{amsmath}
\usepackage{amstext,amsfonts,amsbsy,eucal,amssymb}

\numberwithin{equation}{section}

\def\openone{\leavevmode\hbox{\small1\kern-3.8pt\normalsize1}}

\DeclareMathOperator{\Det}{Det}
\DeclareMathOperator{\Tr}{Tr}
\DeclareMathOperator{\diag}{diag}

\begin{document}

\baselineskip 21pt
\parskip 7pt

%
\noindent
  {\LARGE{
      The extended Lotka-Volterra lattice  \\[3mm]
      and affine Jacobi varieties of spectral curves
      }
    }
\\[5mm]
%
%
\begin{large}
Rei \textsc{Inoue}
  \footnote[2]{E-mail:
    \texttt{rei@gokutan.c.u-tokyo.ac.jp}
    }
\end{large}
\\[3mm]
%
%
  \textsl{Institute of Physics, Graduate School of Arts and Sciences,
    University of Tokyo \\ 
    Komaba 3--8--1, Meguro, Tokyo 153-8902, Japan.
    }
\\[3mm]
%
%
\baselineskip 15pt
\parskip 7pt
%
%
\begin{small}
\textbf{Abstract:} ~
Based on the work by Smirnov and Zeitlin,
we study a simple realization of the
matrix construction of the affine Jacobi varieties.
We find that the realization is given by
a classical integrable model, the extended Lotka-Volterra lattice.
We investigate the integrable structure of the representative
for the gauge  equivalence class of matrices, which is
isomorphic to the affine Jacobi variety,
and make use it to discuss the solvability of the model.
\\[5mm]
%
%
%
\end{small}

\baselineskip 15pt
\parskip 2pt


\section{Introduction}

Consider an $N$ by $N$ matrix whose matrix elements are
polynomials of $z$ of degree $M \in \mathbb{Z}_{>0}$.
We write the characteristic equation of the matrix as
\begin{align}
  \label{curve-general}
  F(z,w)
  \equiv
  w^N - f_1(z) w^{N-1} + f_2(z) w^{N-2} - \cdots
  + (-1)^N f_N(z) = 0,
\end{align}
then each $f_i(z)$ satisfies $\text{deg} f_i(z) \leq iM$.
We assume that the algebraic curve $X$ defined by
\eqref{curve-general} is smooth.
The genus of the curve $X$ is $g = \frac{1}{2} (N-1)(MN-2)$.

For the matrix and the curve $X$,
Beauville introduced an isomorphism \cite{Beauville90},
$$
  \boldsymbol{\mathcal{M}}_F  \simeq X(g) - D.
$$
Here
the left hand side is
the gauge equivalence class $\boldsymbol{\mathcal{M}}_F$ defined as
\begin{align*}
  \boldsymbol{\mathcal{M}}_F
   =
  \{ \mathbf{M}(z) ~|~
      &\deg(\mathbf{M}(z)_{i,j}) \leq M \text{ for all } i,j, ~
      \\
      &\Det | w \openone - \mathbf{M}(z) | = F(z,w) \}
   ~/~ \mathbf{GL}_N(\mathbb{C}),
\end{align*}
and in the right hand side
we have
the set of nontrivial divisors $X(g) = X^g / \mathfrak{S}_g \subset
\mathrm{Div}(X)$
where $\mathfrak{S}_g$ is the symmetric group.
The last term $D$ is a subset of $X(g)$,
where by the Abel transformation
$D$ is mapped to a $(g-1)$-dimensional subvariety of the Jacobi variety
$J(X)$, which is  called the theta divisor $\Theta$.
The Abel transformation induces an isomorphism,
$$
X(g) - D \simeq J(X) - \Theta,
$$
and we call $J(X) - \Theta$ the affine Jacobi variety.
In other words,
the gauge equivalence class $\boldsymbol{\mathcal{M}}_F$
gives a matrix construction of the affine Jacobi variety.
Before Beauville's work,
Mumford studied the case that the curve $X$ is
a hyperelliptic curve (the $N=2$ case),
and introduced a unique representative of the gauge equivalence class
\cite{Mumford-book}.

As discussed in refs.~\citen{Beauville90, Mumford-book,
AdamsHarnadHurtubise90},
the above correspondence of matrices and Jacobi varieties
closely relates to the study of
finite dimensional integrable systems.
The coefficients of the characteristic equation \eqref{curve-general}
correspond to
commuting integrals of motion which
generate $g$ independent vector fields on the affine Jacobi variety.
These fields determine the time-evolution of the divisors in $X(g) - D$,
which is linearized on the Jacobi variety $J(X)$.
Recently Nakayashiki and Smirnov studied
Mumford's representative of a $2$ by $2$ matrix
from the view point of the affine ring for the affine Jacobi variety
\cite{SmirnovNakayashiki00}.
They investigated how the commuting integrals
act on the polynomial ring generated by the matrix elements,
by calculating the cohomology group.
The extension of their work to that for
integrable $N$ by $N$ matrices
is studied by Smirnov and Zeitlin
\cite{Smirnov-Zeitlin0111,Smirnov-Zeitlin0203}.
Starting with an $N$ by $N$ integrable monodromy matrix,
they introduced a unique representative
for the gauge equivalence class $\boldsymbol{\mathcal{M}}_F$
which is isomorphic to a divisor in $X(g) - D$.
The generalization of Mumford's representative appears in 
refs.~\citen{DonagiMarkman96,Smirnov-Zeitlin0203}.
We should remark that
the way of constructing the divisor from the monodromy matrix
is nothing but the separation of variables(SoV) invented by Sklyanin
\cite{Sklyanin95}.

The aim of this paper is
to study a simple realization of the representative for
$\boldsymbol{\mathcal{M}}_F$,
based on ref.~\citen{Smirnov-Zeitlin0111}.
For the characteristic equation \eqref{curve-general}
we have assumed
$$
  f_k(z) = f_k^{(0)} z^{kM} + f_k^{(1)} z^{kM-1} + \cdots + f_k^{(kM)},
   ~~ \text{for } k = 1, \cdots, N,
$$
and we add a condition
\begin{align}
  \label{X-f}
  f_N^{(0)} = 0.
\end{align}
The realization of the representative is given by
the extended Lotka-Volterra lattice
(In some papers we call it the Bogoyavlensky lattice).
This is a classical integrable dynamical model defined by ($1+1$)-dimensional
differential-difference equation \cite{Bogo88,Itoh87,InoueHikami98-Bogo},
\begin{equation}
  \label{Bogo}
  \frac{\mathrm{d} V_n}{\mathrm{d} t}
  =
  2 \,V_n \,
  \sum_{k=1}^{N-1} \left(V_{n+k} - V_{n-k} \right),
\end{equation}
where $V_n \equiv V_n(t), ~n \in \mathbb{Z}$.
In this article we denote this model using LV($N$).
The integrable structure of LV($N$) is
based on
the Poisson algebra $\mathcal{A}_{LV}$ generated by $V_n$,
and on
the $N$ by $N$ Lax matrix given by \cite{Bogo88,InoueHikami98-Bogo}
\begin{equation}
  \label{Bogo-Lax}
  \tilde{\mathbf{L}}_n(z)
    =
  (V_n)^{-\frac{1}{N}}
  \Bigl(
  z^{\frac{1}{N}} \mathbf{E}_{1,1} + (-1)^{N-1} V_n \, \mathbf{E}_{1,N}
  + \sum_{k=1}^{N-1} \mathbf{E}_{k+1,k}
  \Bigr).
\end{equation}
Here $z \in \mathbb{C}$ is a spectral parameter and
$(\mathbf{E}_{i,j})_{m,n} = \delta_{m,i}\delta_{n,j}$.
The Lax matrix composes the monodromy matrix
which generates a family of commuting integrals of motion.
This proves the integrability of the model in Liouville's sense
\cite{InoueHikami98-Bogo}.
It is remarkable that the model has an integrable quantization
which can be applied to
construct the vertex model linked on the crystal base theory
\cite{HikamiInoueKomori99}.
In this paper, we study
LV($N$) of a periodic boundary condition,
which
gives a realization of the representative of
$\boldsymbol{\mathcal{M}}_F$ introduced in ref.~\citen{Smirnov-Zeitlin0111}.
For the case of $N=2$ our realization is essentially same
as that introduced in ref.~\citen{Smirnov0001},
and the previous article gives what corresponds to its generalization.
Based on some special properties of the correspondence of
LV($N$) and the representative,
we try to solve LV($N$) by describing the dynamical variables
$V_n$ \eqref{Bogo} in terms of
the divisor in $X(g) - D$.

The plan of this paper is as follows;
in \S 2, starting with LV($N$)
we construct a monodromy matrix $\overline{\mathbf{T}}(z)$
whose matrix elements
have a special form of polynomials of $z$,
and whose characteristic equation is \eqref{curve-general} with
\eqref{X-f}.
By applying the method SoV
we obtain a divisor in $X(g) - D$.
In \S 3,
following ref.~\citen{Smirnov-Zeitlin0111}
we review the gauge transformation
which derives the representative of $\boldsymbol{\mathcal{M}}_F$,
$\mathbf{M}_F(z)$,
from $\overline{\mathbf{T}}(z)$.
We let $\mathcal{A}_{M_F}$ be the Poisson algebra
for the polynomial ring generated by the coefficients of
the matrix elements of the representative $\mathbf{M}_F(z)$.
Next we investigate some nice properties of
this gauge transformation;
the transformation does not change
the zeros of the separating equation \eqref{B-poly},
and erases the zero mode $B_0$.
These assure the injection from a divisor to
the representative.
After eliminating the center of $\mathcal{A}_{M_F}$,
a family of non-trivial integrals for $\mathcal{A}_{M_F}$ is
composed of $g$ independent variables.
These integrals govern
the evolution of the divisor, which is linearized on $J(X)$.
In \S 4, we investigate LV($N$) by making use of the
results in \S 3.
We show that
the center of $\mathcal{A}_{M_F}$ is a subset of
the center for $\mathcal{A}_{LV}$,
and the non-trivial integrals of motion of both algebras
coincide.
Our claims are that
the representative $\mathbf{M}_F(z)$
can be written in terms of the dynamical variable $V_n$,
and that the structure of $\mathbf{M}_F(z)$
has a close relationship with the solvability
of LV($N$).
The last section \S 5 is devoted to summary and remarks.
We mention the quantization of LV($N$) and
propose some future problems.


\section{Spectral curve of LV($N$) and divisor}

\subsection{Derivation of a proper monodromy matrix}

We study the integrable structure of LV($N$),
and derive a monodromy matrix of a special form which fits to the construction
in ref.~\citen{Smirnov-Zeitlin0111}.

We consider the dynamical system \eqref{Bogo} 
with a periodic boundary condition $V_{n + L} = V_n$, 
and set $L=N(N-1)M, ~ M \in \mathbb{Z}_{>0}$
for later convenience.
The Hamiltonian structure of LV($N$) is defined by the Poisson brackets
\begin{equation}
  \label{M_LV}
  \{ V_n, V_m \}
  =
  2 \,V_m V_n \sum_{k=1}^{N-1} ( \delta_{m,n+k} - \delta_{m,n-k} ),
\end{equation}
and the Hamiltonian $H_1 = \sum_{n=1}^L V_n$ \cite{Bogo88}.
We let $\mathcal{A}_{LV}$ be the Poisson algebra for
$\mathbb{C}[V_n,V_n^{-1}; n \in \mathbb{Z}]$,
whose defining relations are given by \eqref{M_LV}.
After a variable transformation \cite{HikamiInoueKomori99}
\begin{equation}
  \label{V-trans}
  V_n = (P_n P_{n+1} \cdots P_{n+N-1})^{-1} Q_n^{-1} Q_{n+N-1},
\end{equation}
\eqref{M_LV} is transformed into
the Poisson brackets
\begin{align}
  \label{VoltPoisson}
  \{ P_n ~,~ Q_m \}
  = \delta_{n,m} P_n \, Q_n  ,
  ~~~~
  \{ P_n ~,~ P_m \}
  =
  \{ Q_n ~,~ Q_m \} = 0,
\end{align}
where $P_n$ and $Q_n$ are canonical variables.
Using these variables
we apply a gauge transformation to the Lax matrix \eqref{Bogo-Lax},
$
  \mathbf{L}_n(z)
  =
  \boldsymbol{\Omega}_{n+1}(z) \tilde{\mathbf{L}}_n(z)
  \boldsymbol{\Omega}_n(z)^{-1},
$
and obtain the following {\it local} Lax matrix;
\begin{equation}
  \label{Lax-local}
  \mathbf{L}_n(z)
   =
  z^{\frac{1}{N}}
  \Bigl( P_n \mathbf{E}_{1,1} + Q_n \mathbf{E}_{1,2}
  + \frac{1}{z} (-1)^{N-1} Q_n^{-1} \mathbf{E}_{N,1}
  + \sum_{k=2}^{N-1} \mathbf{E}_{k,k+1}
  \Bigr).
\end{equation}
See Appendix A for the concrete description of the
gauge matrix $\boldsymbol{\Omega}_n(z)$.
We introduce another Lax matrix
\begin{align}
  \label{Lax-b}
  \begin{split}
  \overline{\mathbf{L}}_n(z)
   &=
  (\mathbf{L}_n^{-1}(z))^T
   \\
   &=
  \frac{1}{z^{\frac{1}{N}}}
  \Bigl( Q_n^{-1} \mathbf{E}_{1,2}
  + \sum_{k=2}^{N-1} \mathbf{E}_{k,k+1}
  + z (-1)^{N-1} Q_n \mathbf{E}_{N,1}
  + z (-1)^{N-2} P_n \mathbf{E}_{N,2}
  \Bigr),
  \end{split}
\end{align}
where an superscript $^T$ denotes a transposition of the matrices.
Note that $\Det \mathbf{L}_n(z)
= \Det \overline{\mathbf{L}}_n(z) = 1$.
These Lax matrices satisfy the Poisson relations as
\begin{align}
  \label{Poisson-L}
  \begin{split}
  &\{\mathbf{L}_n(z)
  \stackrel{\otimes}{,}
  \mathbf{L}_m(z^\prime) \}
  =
  \delta_{n,m}
  [\, {\mathbf{r}}(z/z^\prime) ~,~ \mathbf{L}_n(z) \otimes
  \mathbf{L}_n(z^\prime) \,],
  \\[1mm]
  &\{\overline{\mathbf{L}}_n(z)
  \stackrel{\otimes}{,}
  \overline{\mathbf{L}}_m(z^\prime) \}
  =
  \delta_{n,m}
  [\, - {\mathbf{r}}(z^\prime/z) ~,~ \overline{\mathbf{L}}_n(z) \otimes
  \overline{\mathbf{L}}_n(z^\prime) \,],
  \\[1mm]
  &\{\mathbf{L}_n(z)
  \stackrel{\otimes}{,}
  \overline{\mathbf{L}}_m(z^\prime) \}
  =
  \delta_{n,m}
  [\, - \mathbf{r}^{T_2}(z/z^\prime) ~,~ \mathbf{L}_n(z) \otimes
  \overline{\mathbf{L}}_n(z^\prime) \,],
  \end{split}
\end{align}
where $\mathbf{r}(z)$ is a classical $r$-matrix;
\begin{align*}
  &{\mathbf{r}}(z)
  = \frac{z + 1}{z - 1}
    \sum_{k=1}^N \mathbf{E}_{k,k} \otimes \mathbf{E}_{k,k}
      +
    \frac{2}{z-1}
    \sum_{1 \leq j < k \leq N}
    \Bigl( \mathbf{E}_{k,j} \otimes \mathbf{E}_{j,k}
           + z \, \mathbf{E}_{j,k} \otimes \mathbf{E}_{k,j}
    \Bigr).
\end{align*}
In deriving the second Poisson relation in \eqref{Poisson-L}, we have used
$\mathbf{r}(z/z^\prime)^{T_1 T_2} = - \mathbf{r}(z^\prime/z)$,
where $^{T_i}$ denotes a transposition in the
$i$-th space.
Now the meaning of the {\it local} Lax matrix becomes clear that
the Lax matrices \eqref{Lax-local} and \eqref{Lax-b}
satisfy the Poisson relations with
$\delta_{n,m}$.

We define two monodromy matrices,
\begin{equation*}
  \mathbf{T}(z) = \prod_{k=1}^{\stackrel{\curvearrowleft}{L}}
                        \mathbf{L}_k(z).
  ~~~~
   \overline{\mathbf{T}}(z) =
    \prod_{k=1}^{\stackrel{\curvearrowleft}{L}}
                        \overline{\mathbf{L}}_k(z).
\end{equation*}
Due to \eqref{Poisson-L}
the monodromy matrices satisfy the following Poisson relations
\begin{align}
  &\{\mathbf{T}(z)
  \stackrel{\otimes}{,}
  \mathbf{T}(z^\prime) \}
  =
  [\, {\mathbf{r}}(z/z^\prime) ~,~ \mathbf{T}(z) \otimes
  \mathbf{T}(z^\prime) \,],
  \\[1mm]
  \label{Poisson-Tbar}
  &\{\overline{\mathbf{T}}(z)
  \stackrel{\otimes}{,}
  \overline{\mathbf{T}}(z^\prime) \}
  =
  [\, - {\mathbf{r}}(z^\prime/z) ~,~ \overline{\mathbf{T}}(z) \otimes
  \overline{\mathbf{T}}(z^\prime) \,],
  \\[1mm]
  &\{{\mathbf{T}}(z)
  \stackrel{\otimes}{,}
  \overline{\mathbf{T}}(z^\prime) \}
  =
  [\, - \mathbf{r}^{T_2}(z/z^\prime) ~,~ \mathbf{T}(z) \otimes
  \overline{\mathbf{T}}(z^\prime) \,].
\end{align}
The first relation denotes that the commuting integrals of motion for
LV($N$) are
generated by $\Tr \mathbf{T}(z)$,
since the Hamiltonian $H_1$ is obtained
by expanding $\Tr \mathbf{T}(z)$ by $z$.
Latter two denote that
the matrix $\overline{\mathbf{T}}(z)$ also generate
the commuting integrals of motion for LV($N$).

The matrix elements of $\mathbf{T}(z)$ and $\overline{\mathbf{T}}(z)$
turn out to be polynomials of $z$,
and these matrices have forms as
\begin{align}
  &\mathbf{T}(z)
  =
  \mathbf{T}_-(z) + \mathbf{T}_0(z) + z \mathbf{T}_+(z),
  \\
  \label{form-Tbar}
  &\overline{\mathbf{T}}(z)
  =
  z \overline{\mathbf{T}}_-(z) + \overline{\mathbf{T}}_0(z) +
   \overline{\mathbf{T}}_+(z).
\end{align}
Here $\mathbf{T}_\pm(z), \overline{\mathbf{T}}_\pm(z)$
are an upper/lower triangular matrices
without diagonal terms, and $\mathbf{T}_0(z),\overline{\mathbf{T}}_0(z)$
are diagonal matrices.
All matrix elements of $\mathbf{T}_\pm(z)$ and $\mathbf{T}_0(z)$ are
polynomials of degree $M(N-1)-1$
but $(\mathbf{T}_0(z))_{1,1}$ which has a polynomial of degree $M(N-1)$.
On the other hand,
elements of $\overline{\mathbf{T}}_0(z)$ are degree $M$
except for $(\overline{\mathbf{T}}_0(z))_{1,1}$ which is degree $M-1$,
and $\overline{\mathbf{T}}_\pm(z)$ has polynomials of degree $M-1$.

We find that the matrix $\overline{\mathbf{T}}(z)$ has
the characteristic equation,
$\Det|w \openone - \overline{\mathbf{T}}(z)| = 0$,
which coincides with
\eqref{curve-general} of \eqref{X-f} and satisfies $f_N(z) = 1$.
Moreover, the construction of the matrix $\overline{\mathbf{T}}(z)$
\eqref{form-Tbar}
and its Poisson relation \eqref{Poisson-Tbar}
are exactly same as those discussed in ref. \citen{Smirnov-Zeitlin0111}
where the Poisson relation
\eqref{Poisson-Tbar}
defines what is called {\it the classical algebra of observables}
generated by the coefficients of polynomials which compose the matrix.
Therefore we conclude that
LV($N$) gives a realization of the algebra of observables.
In the following,
unless we give a notification,
we let $\overline{\mathbf{T}}(z)$ be a matrix
of a special form \eqref{form-Tbar} whose Poisson structure
is given by \eqref{Poisson-Tbar}
and forget about the model LV($N$).


\subsection{Separation of variables and divisor}

We apply SoV method to
obtain the eigenvalues of the monodromy matrix
$\overline{\mathbf{T}}(z)$ algebraically,
following refs. \citen{Sklyanin92, Scott94, Smirnov-Zeitlin0203}.
This method gives a surjective map from
the monodromy matrix $\overline{\mathbf{T}}(z)$ to
a divisor on the curve $X$.

Divide the matrix $\overline{\mathbf{T}}(z)$ into parts as
\begin{align}
  \label{Tbar-devided}
  \overline{\mathbf{T}}(z) =
  \begin{pmatrix}
    a(z) & \vec{b}(z)\\
    \vec{c}(z)^T & \mathbf{d}(z)
  \end{pmatrix},
\end{align}
where
$a(z) = (\overline{\mathbf{T}}(z))_{1,1}$,
$\vec{b}(z)$ and $\vec{c}(z)$ are low
vectors of $N-1$ entries, and $\mathbf{d}(z)$ is an $N-1$ by $N-1$ matrix.
We transform $\overline{\mathbf{T}}(z)$ as
\begin{align*}
  \mathbf{U}(z) = \mathbf{K}\, \overline{\mathbf{T}}(z)\, \mathbf{K}^{-1},
  ~~~
  \mathbf{K} = \openone + \sum_{j=1}^{N-2} k_i \, \mathbf{E}_{i+1,N},
\end{align*}
where $k_i \in \mathbb{C}$.
On the matrix $\mathbf{U}(z)$
we impose some conditions;
\begin{equation}
  \label{diag-condition1}
  (\mathbf{U}(z))_{i,N} = 0, ~ \text{ for } i = 1,\cdots N-1.
\end{equation}
One sees that these conditions reduce to
\begin{align}
  \label{SoV-condition}
    \vec{b}(z) \cdot \vec{x}^T = 0,
    ~~~~
    \vec{x}_i \, \mathbf{d}(z) \cdot \vec{x}^T = 0,
    ~ \text{for} ~ i = 1, \cdots N-2,
\end{align}
where $\vec{x}, ~ \vec{x}_i \in \mathbb{C}^{N-1}$ are low vectors,
$$
  \vec{x}_i= ( 0,\cdots,0, 1, 0,\cdots ,0,k_i ),
  ~~~~
  \vec{x} = (-k_1, -k_2, \cdots, k_{N-2}, 1).
$$
These vectors satisfy $\vec{x}_i \bot \vec{x}$ for all $i$,
then the vectors $\vec{x}_i$ compose basis of the plane normal to $\vec{x}$.
Since the vector $\vec{b}(z)$ is also orthogonal to $\vec{x}$,
it can be uniquely written as
$$
  \vec{b}(z) = \sum_{i=1}^{N-2} \lambda_i \, \vec{x}_i,
  ~~~ \lambda_i \in \mathbb{C}.
$$
By using \eqref{SoV-condition},
we have $\vec{b}(z) \, \mathbf{d}(z) \cdot \vec{x}^T = 0$
which enables to
write $\vec{b}(z) \, \mathbf{d}(z)$ as a linear combination of $\vec{x}_i$
again.
By repeating this procedure, we obtain
$\vec{b}(z) \, \mathbf{d}^k \cdot \vec{x}^T = 0$
for $k \in \mathbb{Z}_{\geq0}$.
Since $\vec{x}$ is not a zero vector,
the condition \eqref{SoV-condition} finally reduces to
\cite{Smirnov-Zeitlin0203}
\begin{align}
  \label{B-eq}
  B(z)
  \equiv
  \Det
  \begin{pmatrix}
    \vec{b}(z) \\
    \vec{b}(z) \mathbf{d}(z) \\
    \vec{b}(z) \mathbf{d}(z)^2  \\
    \vdots \\
    \vec{b}(z) \mathbf{d}(z)^{N-2} \\
  \end{pmatrix}
   = 0.
\end{align}
By the construction of $\overline{\mathbf{T}}(z)$,
$B(z)$ becomes a polynomial of $z$ of degree $g$,
\begin{equation}
  \label{B-poly}
  B(z) = B_0 \prod_{i=1}^g (z - z_i).
\end{equation}
The Poisson relation \eqref{Poisson-Tbar} ensures
that all $z_i$ and $B_0$ are Poisson commutative each other.
For each $z_i$ the eigenvalue of the matrix $\mathbf{U}(z)$,
$w_i \equiv (\mathbf{U}(z_i))_{N,N}$,
is obtained as
\begin{align*}
  w_i
  =
    \Det
    \begin{pmatrix}
      b(z_i) \\
      b(z_i) \mathbf{d}(z_i) \\
      \vdots \\
      b(z_i) \mathbf{d}(z_i)^{N-3} \\
      \vec{\xi} \mathbf{d}(z_i) \\
    \end{pmatrix}
      \Det
     \begin{pmatrix}
      b(z_i) \\
      b(z_i) \mathbf{d}(z_i) \\
      \vdots \\
      b(z_i) \mathbf{d}(z_i)^{N-3} \\
      \vec{\xi} \\
    \end{pmatrix}^{-1},
\end{align*}
where $\vec{\xi}$ is a low vector of $N-1$ entries,
$\vec{\xi} = (0,\cdots, 0,1)$.
The Poisson relation \eqref{Poisson-Tbar}
shows that the separated variables, $w_i$ and $z_i ~ (i = 1, \cdots,g)$,
satisfy the  canonical Poisson brackets;
$$
  \{z_i, z_j\} = \{w_i, w_j\} = 0,
  ~~
  \{z_i , w_j\} = 2 \,\delta_{i,j} z_i \,w_i,
$$
and $B_0$ is a zero mode,
\begin{equation*}
   \{B_0, z_i \} = 0, ~~\{ B_0 , w_i \} = - B_0 w_i.
\end{equation*}

We conclude that
via SoV we get the map from the matrix $\overline{\mathbf{T}}(z)$
to a divisor over $X$, $P = \sum_{i=1}^g [(w_i,z_i)]$,
as each pair of separated variables $(w_i,z_i)$
is a point on the curve $X$.
We assume that
\eqref{B-poly} has different zeros,
$z_i \neq z_j$ for all $i \neq j$,
and that no point $(w_i,z_i)$ coincide with the ramification points
of the map from $X$ to $\mathbb{P}^1$.
These assumptions assure $P \in X(g) - D$.


\section{Integrable monodromy matrix and affine Jacobi variety}

\subsection{Representative of $\boldsymbol{\mathcal{M}}_F$}

Let $\{ \overline{\mathbf{T}}(z) \}_F$ be a set of
matrices with a form \eqref{form-Tbar} and
whose characteristic equations coincide with \eqref{curve-general}.
In the previous section
SoV define an surjective map
from the set $\{ \overline{\mathbf{T}}(z) \}_F$
to a certain set of divisors $P \in X(g) - D$,
but it is not a injective map.
One easily sees the reason by comparing
their dimensions, namely
$\{\overline{\mathbf{T}}(z) \}_F$ and $X(g) - D$ respectively have
$(g+N-1)$-dimension and $g$-dimension as affine spaces.
Smirnov and Zeitlin introduced
a representative of $\boldsymbol{\mathcal{M}}_F$
by setting a gauge transformation
which eliminate the excessive dimension $N-1$ of
$\{\overline{\mathbf{T}}(z)\}_F$.
Following ref.~\citen{Smirnov-Zeitlin0111},
we review the way to introduce the representative of
$\boldsymbol{\mathcal{M}}_F$.

For the matrix $\overline{\mathbf{T}}(z)$
we set
\begin{equation}
  \label{Tbar-mu}
  \overline{\mathbf{T}}(z)
  =
  \boldsymbol{\mu}_0 z^{M} + \boldsymbol{\mu}_1 z^{M-1} +
  \cdots + \boldsymbol{\mu}_M,
\end{equation}
and define $\vec{\nu} = \vec{e_1} \cdot \boldsymbol{\mu_1}$
where $\vec{e_i}$ is a $N$-dimensional low vector
whose entries are zero but $i$-th is $1$.
The gauge transformation from the monodromy matrix $\overline{\mathbf{T}}(z)$
to the representative of $\boldsymbol{\mathcal{M}}_F$,
 $\mathbf{M}_F(z)$, is
\begin{align}
  \label{gauge-S}
  \mathbf{M}_F(z) = \mathbf{S} \, \overline{\mathbf{T}}(z) \, \mathbf{S}^{-1},
   ~~~~ \text{where} ~~
  \mathbf{S} =
  \begin{pmatrix}
    \vec{e}_1 \\
    \vec{\nu} \boldsymbol{\mu}_0^{N-2} \\
    \vdots \\
    \vec{\nu} \boldsymbol{\mu}_0 \\
    \vec{\nu}
  \end{pmatrix}.
\end{align}
Then we obtain $\mathbf{M}_F(z)$ as
\begin{align}
  \label{gauge-M}
  \mathbf{M}_F(z) = \mathbf{U} z^M + O(z^{M-1}),
  ~~~~~
  \mathbf{U}
  =
  \sum_{k=1}^{N}
  m^{(k)} \, \mathbf{E}_{2,k}
  +
  \sum_{k=3}^{N} \mathbf{E}_{k,k-1},
\end{align}
where $m^{(k)}$ are given by
\begin{align*}
  m^{(1)}
  = (-1)^N \Det
   \begin{pmatrix}
    \vec{\nu} \\
    \vec{e}_2 \boldsymbol{\mu}_0 \\
    \vec{e}_3 \boldsymbol{\mu}_0 \\
    \vdots \\
    \vec{e}_N \boldsymbol{\mu}_0 \\
   \end{pmatrix},
  ~~~
  z^{N-1} - \sum_{k=2}^N m^{(k)} z^{N-k}
  =
  \prod_{k=2}^N
    \bigl(z - (\boldsymbol{\mu}_0)_{k,k})\bigr).
\end{align*}
Especially we have
$$
(\mathbf{M}_F(z))_{1,N} = z^{N-1} + O(z^{N-2}),
~~~
(\mathbf{M}_F(z))_{1,i} = O(z^{N-2}),
~\text{for } i = 1, \cdots ,N-1.
$$
The set $\{\overline{\mathbf{T}}(z)\}_F$ is transformed to
$\{\mathbf{M}_F(z)\}$,
and one sees that
$\{\mathbf{M}_F(z)\}$ is a $g$-dimensional affine space.

Under the gauge transformation \eqref{gauge-S},
the zeros of $B(z)$ \eqref{B-poly} are invariant
and the zero mode $B_0$ is canceled
(see Appendix B for the proof);
$$
B(z) \longmapsto {B}_F(z) = (-)^{\frac{1}{2}(N-1)(N-2)}
   \prod_{k=1}^g (z-z_k).
$$
Therefore
a divisor $P = \sum_{i=1}^g [(w_i, z_i)]$
determines ${B}_F(z)$ uniquely.
In conclusion, we get the isomorphism,
$\boldsymbol{\mathcal{M}}_F \simeq X(g) - D$,
where the representative $\mathbf{M}_F(z)$ concretely gives the
matrix construction of the affine Jacobi variety.


\subsection{Integrable system on the Jacobi variety}

Let us see how the integrable structure of the monodromy matrix
$\overline{\mathbf{T}}(z)$ is translated to that of
the matrix $\mathbf{M}_F(z)$.
Via \eqref{gauge-S},
the Poisson structure of
the matrix elements of $\overline{\mathbf{T}}(z)$ \eqref{Poisson-Tbar}
induces the Poisson algebra $\mathcal{A}_{M_F}$ generated by
the matrix elements of $\mathbf{M}_F(z)$.
For the defining relation of $\mathcal{A}_{M_F}$,
see the last part of \S 2 in ref. \citen{Smirnov-Zeitlin0111}
and take its classical limit.
We study a commuting family of integrals of motion for $\mathcal{A}_{M_F}$
without referring the defining relation of $\mathcal{A}_{M_F}$.

From \eqref{Poisson-Tbar}, one obtains
\begin{align}
  \label{center1}
&\{ \Det \overline{\mathbf{T}}(z) \stackrel{\otimes}{,}
\overline{\mathbf{T}}(z^\prime) \} = 0,
\\
  \label{integrals}
&\{ \Det \bigl(w \openone - \overline{\mathbf{T}}(z)\bigr) ,
   \Det \bigl(w^\prime \openone - \overline{\mathbf{T}}(z^\prime)\bigr) \} = 0.
\end{align}
Eq.~\eqref{center1} denotes that
$\Det \overline{\mathbf{T}}(z)$ is Poisson commutative with
all elements of $\overline{\mathbf{T}}(z^\prime)$,
namely $\Det \overline{\mathbf{T}}(z)$ belongs to the
center of $\mathcal{A}_{M_F}$, $\mathcal{A}_{M_F}^0$.
Eq.~\eqref{integrals} assures that the variables
$f_k^{(j)}$ compose
a commutative subalgebra of $\mathcal{A}_{M_F}$,
$\{f_k^{(j)} , f_{k^\prime}^{(j^\prime)}\} = 0$.
Therefore, the dynamical system in $\mathcal{A}_{M_F}$ has
a family of integrals of motion,
$\{f_k^{(j)} ~|~ k = 1, \cdots N-1, ~ j = 0, \cdots kM \}$, whose number is
$g + 2(N-1)$.
In the following we show that $2(N-1)$ integrals,
$f_k^{(0)}$ and $f_k^{(kM)}$, $k=1, \cdots N-1$, belong to
$\mathcal{A}_{M_F}^0$, namely
the number of non-trivial integrals of motion is $g$.
What we should show is
\begin{align}
  \label{center2}
  \{ f_k^{(j)} \stackrel{\otimes}{,}  \mathbf{M}_F(z) \}
  =
  0,  ~~~ \text{for }  k = 1, \cdots, N-1 \text{ and } j = 0, kM.
\end{align}
One sees that
since $f_k(z)$ can be written in terms of
$t_k(z) \equiv \Tr (\overline{\mathbf{T}}(z))^k$;
$$
f_1(z) = t_1(z), ~~ f_2(z) = \frac{1}{2} (t_1(z)^2 - t_2(z)), ~~\cdots
$$
\eqref{center2} are reduced to
\begin{equation}
  \label{Poisson-tM}
  \{ t_k^{(j)} \stackrel{\otimes}{,}  \mathbf{M}_F(z) \}  = 0,
  ~~~ \text{for } j = 0, kM.
\end{equation}
Here we denote
the dominant terms of $\Tr (\overline{\mathbf{T}}(z))^k$
in the $z \to 0,\infty$ limits using $t_k^{(kM)}, ~t_k^{(0)}$ respectively.
The Poisson relation \eqref{Poisson-Tbar} reduces to
\begin{align*}
  \{ t_k(z) \stackrel{\otimes}{,} \overline{\mathbf{T}}(z^\prime) \}
  =
  \Tr_1 \{ (\overline{\mathbf{T}}(z))^k
           \stackrel{\otimes}{,} \overline{\mathbf{T}}(z^\prime) \}
  =
  k \, \Tr_1 [ \,\mathbf{r}(z/z^\prime), (\overline{\mathbf{T}}(z))^k \otimes
                         \overline{\mathbf{T}}(z^\prime)\,],
\end{align*}
which derives
\begin{align}
  \label{Poisson-tTbar}
  \begin{split}
  &\{ t_k^{(j)} \stackrel{\otimes}{,} \overline{\mathbf{T}}(z^\prime) \}
  =
  k \, [(\mathbf{K}^{(j)})^k , \overline{\mathbf{T}}(z^\prime)],
  \\
  &\{ t_k^{(j)} \stackrel{\otimes}{,} \vec\nu \}
  =
  - k \, \vec\nu \,(\mathbf{K}^{(j)})^k,
  \\
  &\{ t_k^{(j)} \stackrel{\otimes}{,} \boldsymbol{\mu}_0 \}
  =
  k \, [\, (\mathbf{K}^{(j)})^k, \boldsymbol{\mu}_0 \,],
  \end{split}
\end{align}
for $j = 0, kM$.
Here we use
the matrices $\boldsymbol{\mu}_k$ \eqref{Tbar-mu}
and
\begin{align*}
  &\mathbf{K}^{(0)} = \diag[0,
                           (\boldsymbol{\mu}_{0})_{2,2},
                           \cdots,(\boldsymbol{\mu}_{0})_{N,N}],
  \\
  &\mathbf{K}^{(kM)} = - \diag[0,
                           (\boldsymbol{\mu}_{M})_{2,2},
                           \cdots,(\boldsymbol{\mu}_{M})_{N,N}].
\end{align*}
Due to the relations \eqref{Poisson-tTbar} we get
\begin{align}
  \label{Poisson-tS}
  \{ t_k^{(i)} \stackrel{\otimes}{,} \mathbf{S} \}
  = - k \, (\mathbf{S} - \mathbf{E}_{1,1} ) (\mathbf{K}^{(i)})^k,
  ~~~ \text{for } i = 0,kM,
\end{align}
and \eqref{Poisson-tM} is proved.

We arrange the non-trivial integrals as
$$
f_1^{(1)},  \cdots, f_1^{(M-1)},f_2^{(1)},\cdots,
f_2^{(2M-1)}, \cdots, f_{N-1}^{(1)}, \cdots, f_{N-1}^{(M(N-1)-1)},
$$
and number them in order,
\begin{equation}
  \label{Hamiltonians}
  \mathcal{H}_1,\mathcal{H}_2, \cdots, \mathcal{H}_g.
\end{equation}
In conclusion, we obtain the integrable structure of $\mathcal{A}_{M_F}$
that the $g$ commuting integrals $\mathcal{H}_i$ describe
the time evolution
for $\mathcal{O} \in \mathcal{A}_{M_F}$;
\begin{equation}
  \label{time-evol}
  \frac{\partial \mathcal {O}}{\partial \tau_i}
  \equiv \{ \mathcal{H}_i , \mathcal{O} \},
  ~~~ \text{for~ } i = 1, \cdots, g.
\end{equation}
On the Jacobi variety $J(X)$, $\mathcal{H}_i$ generate the invariant vector
field
where the time evolution of the image of the divisor $P$
is linearized.
By the inverse map of the Abel transformation,
we get $z_i$ as functions of times $\tau_i$,
$z_i = z_i(\tau_1, \cdots \tau_g)$
once the initial values $z_i(0,\cdots,0) = z_i^0$
is given.
Note that the curve $X$ is determined by
the eigenvalues of $\mathcal{H}_i$.


\section{Description of the LV($N$)}

Now we study the realization of
$\mathbf{M}_F(z)$ given by LV($N$).
Since the characteristic equation \eqref{curve-general} is invariant under
the gauge transformation,
the commuting family can be written in terms of the dynamical variables of
LV($N$)
by making use of the Lax matrix $\Tilde{\mathbf{L}}_n(z)$
\eqref{Bogo-Lax}.
We introduce variables $\mathcal{P}_k$ and $\mathcal{P}^\prime_k$;
\begin{align*}
  &\mathcal{P}_0 = \prod_{n=1}^L (V_n)^{-\frac{1}{N}},
  \\
  &\mathcal{P}_{k} = \prod_{n=1}^{NM} (V_{(N-1)n+k}),
   ~~~ \text{for } k = 1, \cdots, N-1,
  \\
  &\mathcal{P^\prime}_{k} = \prod_{n=1}^{(N-1)M} (V_{Nn+k}),
   ~~~ \text{for } k = 1, \cdots, N.
\end{align*}
These variables constitute the center of $\mathcal{A}_{LV}$, 
$\mathcal{A}_{LV}^0$. 
Note that not all of them are independent, and
the generators of $\mathcal{A}_{LV}^0$ are obtained by
choosing any $2(N-1)$ variables from
$\{\mathcal{P}_{k \geq 1}, \mathcal{P}_k^\prime \}$.
The direct calculations show that
the elements of $\mathcal{A}_{M_F}^0$ are written in terms
of these variables as
\begin{align*}
  &f_N(z) = \Det \overline{\mathbf{T}}(z) = 1,
  \\
  &f_{N-1}^{(0)} = \mathcal{P}_0,
  \\
  &f_1^{(0)} = \mathcal{P}_0^{-1}
               ( \mathcal{P}_1^{-1} + \cdots + \mathcal{P}_{N-1}^{-1} ),
  \\[1mm]
  &f_{N-1}^{(N-1)M} = \mathcal{P}_0
               ( \mathcal{P^\prime}_1 + \cdots + \mathcal{P^\prime}_{N} ),
  \\[1mm]
  &f_1^{(N-1)M} = \mathcal{P}_0^{-1}
               ( \mathcal{P^\prime}_1^{-1} + \cdots +
\mathcal{P^\prime}_{N}^{-1} ),
\end{align*}
and that other elements,
$f^{(0)}_k, ~ f^{(kM)}_k$ for $k = 2, \cdots, N-2$,
are obtained from the above.
Therefore we see $\mathcal{A}_{M_F}^0 \subset \mathcal{A}_{LV}^0$,
and the non-trivial integrals of motion for LV($N$)
have one-to-one correspondence to $\mathcal{H}_i$.

Based on the above observation,
we conjecture that
\\[2mm]
(i)
the matrix $\mathbf{M}_F(z)$ can be written in terms of
$V_n$,
\\
(ii)
then all zeros of $B_F(z)$ are given by $V_n$, $z_i = z_i(\{V_n\})$,
and we can {\it solve} LV($N$) as
$$
V_n = V_n(\mathcal{P}_k, \mathcal{P}_k^\prime, \mathcal{H}_i
; z_i).
$$
To discuss the conjecture, using \eqref{Lax-x} and
\eqref{gauge-S} we rewrite $\mathbf{M}_F(z)$ as
\begin{equation}
  \label{LV-M}
  \mathbf{M}_F(z) =
  \mathbf{S} \, (\boldsymbol{\Omega}_1(z)^{-1})^T
  \Bigl(\bigl(\Tilde{\mathbf{L}}_L(z)
  \cdots \Tilde{\mathbf{L}}_1(z)\bigr)^{-1}\Bigr)^T
  \boldsymbol{\Omega}_1(z)^T \,\mathbf{S}^{-1}.
\end{equation}
Due to the
construction of the gauge matrix $\boldsymbol{\Omega}_1(z)$
\eqref{gauge-Omega},
we reduce \eqref{LV-M} to
\begin{align}
  \label{LV-M-V}
  \mathbf{M}_F(z) =
  \Tilde{\mathbf{S}}\, \mathbf{X}(z)^{-1}
  \Bigl(\bigl(\Tilde{\mathbf{L}}_L(z)
  \cdots \Tilde{\mathbf{L}}_1(z)\bigr)^{-1}\Bigr)^T
  \mathbf{X}(z) \,\Tilde{\mathbf{S}}^{-1}.
\end{align}
We have conjectured by (i)
that the matrix $\Tilde{\mathbf{S}}$ is written
in terms of $V_n$.
Remember that LV($N$)
has $N(N-1)M$ dynamical variables $V_n$.
Once we accept (i), (ii) follows (i)
since we have enough number of
relations
to describe $V_n$ in terms of
$\mathcal{P}_k, \mathcal{P}_k^\prime, \mathcal{H}_i$ and $z_i$.
Actually we have $g$ relation equations between $z_i$ and $V_n$,
$g$ non-trivial integrals of motion $\mathcal{H}_i$,
and $2(N-1)$ independent generators of $\mathcal{A}_{LV}^0$,
whose summation coincides with $N(N-1)M$.
It should be remarked that
due to the periodic boundary condition of the system,
we essentially have the translation invariant
such as
$\mathcal{P}_1 = \cdots = \mathcal{P}_{N-1}$ and  $\mathcal{P}^\prime_1 = \cdots = \mathcal{P}^\prime_{N}$.

In the following, we study the cases of $N=2,3$
which illustrate the correspondence of
LV($N$) and the integrable structure on the affine Jacobi variety.
We prove the conjecture in the $N=2$ case, and the simplest case of $N=3$.
For general $N$, it seems to be very complicated even to show (i).
\\

\noindent
$\bullet$ {$N=2$ case}

We have $L = 2M$, $g = M-1$,
the integrals of motion $\mathcal{H}_i, ~i = 1,\cdots M-1$,
and the elements of $\mathcal{A}_{LV}^0$,
$$
\mathcal{P}_1 = (\mathcal{P}_0)^{-2} =
\prod_{k=1}^{2M} V_k,
~~~
\mathcal{P}_1^\prime = \prod_{k=1}^{M} V_{2k-1},
~~~
\mathcal{P}_2^\prime = \prod_{k=1}^{M} V_{2k}.
$$
By definition,
we have
\begin{align*}
  &\mathbf{S}
  =
  \begin{pmatrix}
    1 & 0 \\
    -P_2 \cdots P_{L-1} Q_1 Q_L^{-1} & P_1 \cdots P_{L-1} Q_L^{-1}
  \end{pmatrix}
  =
  \begin{pmatrix}
    1 & 0 \\
    - \mathcal{P}_0 V_L & \mathcal{P}_0 P_L^{-1} Q_L^{-1}
  \end{pmatrix},
  \\[1mm]
  &\mathbf{B}_1^{-1}
  =
   P_1^{\frac{1}{2}} Q_1^{-\frac{1}{2}}
  \begin{pmatrix}
   1 & 0 \\
   0 & P_1^{-1} Q_1
  \end{pmatrix},
\end{align*}
where $P_n, Q_n$ are canonical variables \eqref{VoltPoisson},
and $V_n = (P_n P_{n+1})^{-1}Q_n^{-1} Q_{n+1}$ \eqref{V-trans}.
Then \eqref{LV-M} reduces to \eqref{LV-M-V} where the matrix
$\Tilde{\mathbf{S}}$ is
\begin{align*}
  \Tilde{\mathbf{S}}
   =
   \begin{pmatrix}
     1 & 1 \\
     0 & - \mathcal{P}_0 V_L
   \end{pmatrix},
\end{align*}
which justifies (i).

Let us consider the case of $M=2$, $L=4$ and $g=1$.
Now \eqref{curve-general} becomes
\begin{equation*}
  w^2 - \mathcal{P}_0 (z^2 - H_1 z +
   \mathcal{P}^\prime_1 +  \mathcal{P}^\prime_2 ) w
  +1 = 0,
\end{equation*}
where we have three independent integrals of motion;
\begin{align}
  \label{N=2-Hamiltonian}
H_1 = \sum_{k=1}^4 V_n,
~~~
& \mathcal{P}^\prime_1 = V_1 V_3, ~~   \mathcal{P}^\prime_2 = V_2 V_4.
\end{align}
Due to the translation invariant we set
$\mathcal{P} \equiv \mathcal{P}^\prime_1 = \mathcal{P}^\prime_2$
which yields
$\mathcal{P}_0^2 = \mathcal{P}^{-2}$.
The polynomial $B_F(z)$ has a zero $z_1$,
\begin{equation}
  \label{N=2-divisor}
  z_1 = V_1 + V_2.
\end{equation}
Here $z_1$ is a function of $\tau_1$
defined by \eqref{time-evol} with $\mathcal{H}_1 = \mathcal{P}_0 H_1$.
Finally, the dynamical variables
$V_n = V_n(\mathcal{P},\mathcal{H}_1; z_1(\tau_1))$ for $n = 1,2,3,4$
are obtained from
\eqref{N=2-Hamiltonian} \eqref{N=2-divisor}.
\\

\noindent
$\bullet$ {$N=3$ case}

We consider the $L=6, M=1$ and $g=1$ case.
Now the characteristic equation is
\begin{align*}
  w^3 & + \mathcal{P}_0^2 \bigl( z (\mathcal{P}_1 +  \mathcal{P}_2)
                          - (\mathcal{P}^{\prime}_1 \mathcal{P}^{\prime}_2
                             + \mathcal{P}^{\prime}_1 \mathcal{P}^{\prime}_3
                             + \mathcal{P}^{\prime}_2 \mathcal{P}^{\prime}_3
                            )\bigr) w^2
   \\
   & \hspace{3cm}+ \mathcal{P}_0 (z^2 + z H_1 +
                   \mathcal{P}^{\prime}_1 + \mathcal{P}^{\prime}_2 +
                   \mathcal{P}^{\prime}_3) w
   - 1
   = 0,
\end{align*}
where
\begin{align}
  \label{N=3-Hamiltonian}
  \mathcal{P}_i = V_i V_{i+2} V_{i+4}, ~ \text{for } i = 1,2,
  ~~~
  \mathcal{P}^\prime_i = V_i V_{i+3}, ~ \text{for } i = 1,2,3,
  ~~~
  H_1 = \sum_{k=1}^6 V_n.
\end{align}
We set $\mathcal{P}^\prime_i \equiv \mathcal{P}$,
and for simplicity consider the case of
$\mathcal{P}_i \equiv \mathcal{P}^{\frac{3}{2}}$
and $\mathcal{P}_0 \equiv \mathcal{P}^{-1}$.
Then the matrix $\mathbf{M}(z)$ reduces to \eqref{LV-M-V}
where the gauge matrix $\Tilde{\mathbf{S}}$ is written in terms of $V_n$,
\begin{align*}
  \Tilde{\mathbf{S}}
  =
  \begin{pmatrix}
   0  & 1 & 1 \\
   \mathcal{P}^{-1} (V_5 + V_6) + 2 \mathcal{P}^{-\frac{1}{2}} &
   \mathcal{P}^{-1} V_6 &
   - \mathcal{P}^{-1} (V_5 + V_6) - 2 \mathcal{P}^{-\frac{1}{2}} \\
   -1 & - \mathcal{P}^{-\frac{1}{2}}  V_6 & 1
  \end{pmatrix}.
\end{align*}
The polynomial $B_F(z)$ has a zero $z_1$,
\begin{equation}
  \label{N=3-divisor}
  z_1 = - (V_1 + V_2) \Bigl( \frac{V_3 V_4}
                            {\mathcal{P}^{\frac{1}{2}} (V_3 + V_4)
                              + \mathcal{P}}
                        + 1 \Bigr).
\end{equation}
As same as the $N=2$ case,
we obtain $V_n = V_n(\mathcal{P},\mathcal{H}_1;z_1(\tau_1))$
by using \eqref{N=3-Hamiltonian} and \eqref{N=3-divisor}.
\\[1mm]

For the general $N$ cases
we support (i) and (ii), and
the dynamical variables of LV($N$)
should be solved as
$$
V_n = V_n(\mathcal{P},\mathcal{H}_1, \cdots, \mathcal{H}_g;
           z_1, \cdots , z_g),
$$
where $z_i = z_i(\tau_1, \cdots, \tau_g)$.


\section{Summary and remarks}

In this paper,
we have studied the realization of the
representative of the gauge equivalence class $\boldsymbol{\mathcal{M}}_F$,
which is given by the classical integrable model,
the extended Lotka-Volterra lattice.

The gauge equivalence class 
$\boldsymbol{\mathcal{M}}_F$ have the elements 
whose characteristic equation \eqref{curve-general} is common,
and 
the coefficients of \eqref{curve-general} correspond to
a set of commuting integrals of motion.
There is the isomorphism from $\boldsymbol{\mathcal{M}}_F$ to
a set of divisors $X(g) - D$,
and the time-evolution of the divisor
is linearized on the Jacobi variety $J(X)$.
In ref.~\citen{Smirnov-Zeitlin0111},
it was introduced that
the way to construct the representative of $\boldsymbol{\mathcal{M}}_F$
by starting with the integrable monodromy matrix
$\overline{\mathbf{T}}(z)$ \eqref{form-Tbar}.
Based on the integrable Poisson structure of the monodromy matrix,
the divisor is determined via SoV.

We have found that
LV($N$) gives the realizations
not only for the monodromy matrix $\overline{\mathbf{T}}(z)$ but also for the
representative $\mathbf{M}_F(z)$.
We have studied the correspondence of
LV($N$) and
the representative and their Poisson algebras in detail.
Then we have shown that
the family of non-trivial integrals of motion for
the representative coincides with
that of LV($N$), and that the number of these integrals are
necessary and sufficient to describe the model.
Especially our claim is that
$\mathbf{M}_F(z)$ can be written in terms of the dynamical variables
of LV($N$). 
These make possible to solve the model, and
the time evolutions of the dynamical variables
$V_n$ are obtained as
$$
V_n = V_n(\mathcal{P},\mathcal{H}_1, \cdots, \mathcal{H}_g;
           z_1(\{ \tau_i \}), \cdots , z_g(\{ \tau_i \})).
$$
Here $\mathcal{P}$ and $\mathcal{H}_i$ are the integrals of motion,
where each $\mathcal{H}_i$ generates the independent time
$\tau_i$. 

In closing,
we would like to mention the
quantization of LV($N$).
By replacing the canonical variables \eqref{VoltPoisson}
in the Lax matrices $\mathbf{L}_n(z)$ \eqref{Lax-local}
and $\overline{\mathbf{L}}_n(z)$ \eqref{Lax-b}
with the Weyl operators,
\begin{align*}
  [ \hat{P}_n, \hat{Q}_m ] = \delta_{n,m} \hat{P}_n \hat{Q}_n,
  ~~~~
  [ \hat{P}_n, \hat{P}_m ] = [\hat{Q}_n , \hat{Q}_m ] = 0,
\end{align*}
we get the quantum integrable model
\cite{InoueHikami98-Bogo,HikamiInoueKomori99}.
As same as the classical case
we note the Lax matrix $\overline{\mathbf{L}}_n(z)$ \eqref{Lax-b}.
Now this matrix
with Weyl operators satisfies
the fundamental commuting relation
\begin{align*}
  \mathbf{R}(z/z^\prime ; q) \,
  \bigl(\overline{\mathbf{L}}_n(z) \otimes \openone\bigr)
  \bigl(\openone \otimes  \overline{\mathbf{L}}_n (z^\prime)\bigr)
  =
  \bigl(\openone \otimes  \overline{\mathbf{L}}_n (z^\prime)\bigr)
  \bigl(\overline{\mathbf{L}}_n(z) \otimes \openone\bigr) \,
  \mathbf{R}(z/z^\prime ; q),
\end{align*}
where $R$-matrix is
\begin{align*}
  \mathbf{R}(z ;q)
   &=
  \sum_{k=1}^{N} \, (z - q^2) \,
  \mathbf{E}_{k,k} \otimes \mathbf{E}_{k,k}
  +
  \sum_{j=1}^{N} \, \sum_{k=1}^{N-1} \,
  q \, (z - 1) \,
  \mathbf{E}_{j,j} \otimes \mathbf{E}_{j+k,j+k}
  \\ \nonumber
  & ~~~~~~~~
  +
  \sum_{1 \leq j < k \leq N}
  (1 - q^2) \,
  \bigl(
   \, \mathbf{E}_{j,k} \otimes \mathbf{E}_{k,j}
  +
  z \mathbf{E}_{k,j} \otimes \mathbf{E}_{j,k}
  \bigr).
\end{align*}
Especially in the $N=2$ case the Lax matrix becomes
\begin{align}
   \label{LbN=2}
    \overline{\mathbf{L}}_n(z)
    =
   \frac{1}{z^{\frac{1}{2}}}
  \begin{pmatrix}
    0 & \hat{Q}_n^{~-1} \\
    - z \hat{Q}_n & z \hat{P}_n
  \end{pmatrix}.
\end{align}
Taking into account the canonical transformation,
this is essentially same as what discussed in ref. \citen{Smirnov0001}.
In general $N$ case,
the monodromy matrix $\overline{\mathbf{T}}(z)$
is written as
\begin{align*}
  \overline{\mathbf{T}}(z)
   &=
   \prod_{k=1}^{\stackrel{\curvearrowleft}{MN}}
   \mathbb{L}_k(z),
\end{align*}
where
$
   \mathbb{L}_k(z)
    =
   \Bigl( \overline{\mathbf{L}}_{k(N-1)}(z)
           \overline{\mathbf{L}}_{k(N-1)-1}(z)
           \cdots
           \overline{\mathbf{L}}_{(k-1)(N-1)+1}(z)
   \Bigr)
$
is
\begin{align*}
   \frac{1}{z^{\frac{N-1}{N}}}
   \begin{pmatrix}
     0 & 0 & \cdots & 0 & \hat{Q}_{N-1}^{-1} \\
     z (-)^{N-1}\hat{Q}_1 & z (-)^{N-2}\hat{P}_1 & 0 & \cdots &  0 \\
     0 & z (-)^{N-1}\hat{Q}_2 \hat{Q}_1^{-1} & z (-)^{N-2} \hat{P}_2 & 0
&  \vdots \\
     \vdots & 0&  \ddots & \ddots & 0\\
     0 & \cdots & 0 & z (-)^{N-1}\hat{Q}_{N-1}\hat{Q}_{N-2}^{-1} &
z (-)^{N-2}\hat{P}_{N-1}
   \end{pmatrix}_{[k-1]}.
\end{align*}
Here in the matrix with the subscript $_{[k-1]}$ 
the operators $\hat{P}_i$ and $\hat{Q}_i$ are regarded 
as $\hat{P}_{i+(k-1)(N-1)}$ and $\hat{Q}_{i+(k-1)(N-1)}$ 
respectively.            
We expect that
the matrix $\mathbb{L}_n(z)$ gives a key to generalize
the Baxter equations and their dual structure
in the sense of ref.~\citen{Smirnov0001},
which is a future problem.
The matrix $\mathbb{L}_k(z)$
may link to the relatives
or the extension to $\mathfrak{sl}_N$ of the Toda lattice.
It seems to be interesting to study this matrix
in both of classical and quantum cases from this point of view.


\subsection*{Acknowledgements}

The author thanks Prof.~A. Nakayashiki for helpful comments,
and Prof.~A. Kuniba for kind encouragement.
She appreciates discussions with T. Yamazaki.
She is grateful to Prof. B. de Wit and members of ITP and Spinoza Institute
in Utrecht University for their hospitality.
A part of this work was done during her stay at Spinoza institute.
R. I. is a Research Fellow of the Japan Society for the Promotion of Science.


\newpage

\setcounter{equation}{0}

\renewcommand{\theequation}{A.\arabic{equation}}

\addcontentsline{toc}{chapter}{Appendix \protect\numberline{A}
                              {Local Lax matrix for LV($N$)}}

\subsection*{Appendix A ~ Local Lax matrix for LV($N$)}

We define a gauge transformation of $\tilde{\mathbf{L}}(x)$ \eqref{Bogo-Lax}
as follows;
\begin{align}
  \label{Lax-x}
  \mathbf{L}_n(z)
  =
  \boldsymbol{\Omega}_{n+1}(z) \tilde{\mathbf{L}}_n(z)
  \boldsymbol{\Omega}_n(z)^{-1},
   ~~~~
\end{align}
Here the gauge matrix $\boldsymbol{\Omega}_n(z)$ is
\begin{equation}
  \label{gauge-Omega}
  \boldsymbol{\Omega}_n(z) =
   \mathbf{B}_n \, \mathbf{A} \, \mathbf{X}(z),
\end{equation}
where
\begin{align*}
 &\mathbf{A}
 = \Bigl(\sum_{k=1}^{N} \, \mathbf{E}_{k,N+1-k} \Bigr)
    \Bigl(\openone \, - \, \sum_{k=1}^{N-1} \,
                                \mathbf{E}_{k,k+1} \Bigr),
 \nonumber \\
 &\mathbf{B}_n
 = \prod_{k=0}^{N-2}\,
    \Bigl( P_{n+k} \Bigr)^{\displaystyle
      - \sum_{j=k+2}^{N} \mathbf{D}^{(j)}
                    }
    \Bigl(   Q_{n+k} \Bigr)^{\displaystyle \mathbf{D}^{(k+2)}} ,
 \nonumber \\
 &\mathbf{X}(z)
 = \sum_{k=1}^{N} \, z^{\frac{k-1}{N}} \mathbf{E}_{k,k}
\end{align*}
and we use
\begin{equation*}
 \mathbf{D}^{(j)}
 = \frac{1}{N} \, \openone \, - \, \mathbf{E}_{j,j},
  ~~~~~
   \Bigl(P_{n} \Bigr)^{\displaystyle \mathbf{D}^{(j)}}
    =
   \sum_{k=1}^N P_n^{(\mathbf{D}^{(j)})_{k,k}} \mathbf{E}_{k,k}.
\end{equation*}
In the above we have used a notation,
\begin{align*}
  P_n^{\,\displaystyle \mathbf{D}}
  =
  \diag[P_n^{\,d_1}, P_n^{\,d_2}, \cdots, P_n^{\,d_N}],
  ~~ \text{where} ~~
  \mathbf{D} = \diag[d_1, \cdots, d_N].
\end{align*}
Finally we obtain the local Lax matrix \eqref{Lax-b},
\begin{equation*}
  \mathbf{L}_n(z)
   =
  z^{\frac{1}{N}}
  \Bigl( P_n \mathbf{E}_{1,1} + Q_n \mathbf{E}_{1,2}
  + \frac{1}{z} (-1)^{N-1} Q_n^{-1} \mathbf{E}_{N,1}
  + \sum_{k=2}^{N-1} \mathbf{E}_{k,k+1}
  \Bigr).
\end{equation*}
Note that the gauge matrix $\boldsymbol{\Omega}_n(z)$ is
different from that introduced in ref.~\citen{HikamiInoueKomori99}.


\setcounter{equation}{0}

\renewcommand{\theequation}{B.\arabic{equation}}

\addcontentsline{toc}{chapter}{Appendix \protect\numberline{B}
                              {Gauge transformation of $B(z)$}}

\subsection*{Appendix B ~ Gauge transformation of $B(z)$}

We divide the matrix $\mathbf{S}$ \eqref{gauge-S}
in the same way as \eqref{Tbar-devided},
\begin{align*}
  \mathbf{S} =
  \begin{pmatrix}
    1 & \vec{0} \\
    \vec{s}_1^{\,T} & \mathbf{s}_2
  \end{pmatrix},
\end{align*}
where we use
\begin{align*}
  \vec{s}_1 = \vec{c}_0 \,(\mathbf{d}_0^{N-1-i})^T,
  ~~~~
  \mathbf{s}_2 =
  \begin{pmatrix}
    \vec{b}_0 \mathbf{d}_0^{N-2} \\
    \vdots \\
    \vec{b}_0 \mathbf{d}_0 \\
    \vec{b}_0
  \end{pmatrix},
\end{align*}
and $\vec{c}_0, \vec{b}_0$ and $\mathbf{d}_0$ are
dominant parts of $\vec{c}(z), \vec{b}(z)$ and $\mathbf{d}(z)$
in $z \to \infty$.
The matrix $\mathbf{S}$
transforms the monodromy matrix $\overline{\mathbf{T}}(z)$ to
\begin{align*}
  \mathbf{M}_F(z)
  = \mathbf{S}\, \overline{\mathbf{T}}(z) \,\mathbf{S}^{-1}
  =
  \begin{pmatrix}
    \sharp & \vec{b}(z)\, \mathbf{s}_2^{-1} \\
    \flat & \bigl({\vec{s}_1}^{\,T} \vec{b}(z) +
                  \mathbf{s}_2 \,\mathbf{d}(z) \bigr) \,\mathbf{s}_2^{-1}
  \end{pmatrix},
\end{align*}
where the parts indicated by $\sharp$ and $\flat$ are not important now.
Following this transformation,
the polynomial $B(z)$ \eqref{B-poly} becomes $B_F(z)$ as
\begin{align*}
  B_F(z)
  &=
  \Det
  \begin{pmatrix}
    \vec{b}(z)\, \mathbf{s}_2^{-1} \\
    \vec{b}(z)\, \mathbf{s}_2^{-1}
       \bigl({\vec{s}_1}^{\,T} \vec{b}(z) \mathbf{s}_2^{-1} +
                  \mathbf{s}_2 \, \mathbf{d}(z) \,\mathbf{s}_2^{-1} \bigr) \\
    \vdots \\
    \vec{b}(z)\, \mathbf{s}_2^{-1}
       \bigl({\vec{s}_1}^{\,T} \vec{b}(z) \mathbf{s}_2^{-1} +
                  \mathbf{s}_2 \, \mathbf{d}(z) \,\mathbf{s}_2^{-1} \bigr)^{N-2}
\\
  \end{pmatrix}
  \\
  &=
  \Det
  \begin{pmatrix}
    \vec{b}(z) \\
    \vec{b}(z) \mathbf{d}(z) \\
    \vdots \\
    \vec{b}(z) \mathbf{d}(z)^{N-2} \\
  \end{pmatrix}
  \Det (\mathbf{s}_2^{-1})
  \\
  &= (-)^{\frac{1}{2}(N-1)(N-2)} B(z) \,B_0^{-1}
\end{align*}
The second equality is due to
$$
  \vec{b}(z) \mathbf{s}_2^{-1} \vec{s}_1^{\,T} \vec{b}(z)
  ~\propto~ \vec{b}(z),
$$
and the third one follows \eqref{B-eq} and
$$
\Det \mathbf{s}_2 = (-)^{\frac{1}{2}(N-1)(N-2)} B_0
$$
where $B_0$ is the zero mode of $B(z)$.
Finally we obtain $B_F(z)$ which do not have the zero mode.



\begin{thebibliography}{10}

\bibitem{Beauville90}
A.~Beauville, Acta. Math. \textbf{164}, 211 (1990).

\bibitem{Mumford-book}
D.~Mumford, {\it Tata Lectures on Theta II\/} (Birkh\"auser, 1984).

\bibitem{AdamsHarnadHurtubise90}
M.~R. Adams, J.~Harnad, and J.~Hurtubise, Comm. Math. Phys. \textbf{134}, 555
  (1990).

\bibitem{SmirnovNakayashiki00}
A.~Nakayashiki and F.~A. Smirnov, Comm. Math. Phys. \textbf{217}, 623 (2001),
  math-ph/0001017.

\bibitem{Smirnov-Zeitlin0111}
F.~A. Smirnov and V.~Zeitlin, math-ph/0111038.

\bibitem{Smirnov-Zeitlin0203}
F.~A. Smirnov and V.~Zeitlin, math-ph/0203037.

\bibitem{DonagiMarkman96}
R.~Donagi and E.~Markman, Lecture Notes in Mathematics \textbf{1620}, 1 (1996).

\bibitem{Sklyanin95}
E.~K. Sklyanin, Prog. Theor. Phys. Suppl. \textbf{118}, 35 (1995).

\bibitem{Bogo88}
O.~I. Bogoyavlensky, Phys. Lett. A \textbf{134}, 34 (1988).

\bibitem{Itoh87}
Y.~Itoh, Prog. Theor. Phys. \textbf{78}, 507 (1987).

\bibitem{InoueHikami98-Bogo}
R.~Inoue and K.~Hikami, J. Phys. Soc. Jpn. \textbf{67}, 87 (1998).

\bibitem{HikamiInoueKomori99}
K.~Hikami, R.~Inoue, and Y.~Komori, J. Phys. Soc. Jpn. \textbf{68}, 2234
  (1999).

\bibitem{Smirnov0001}
F.~A. Smirnov, J. Phys. A: Math. Gen. \textbf{33}, 3385 (2000),
  math-ph/0001032.

\bibitem{Sklyanin92}
E.~K. Sklyanin, Comm. Math. Phys. \textbf{150}, 181 (1992).

\bibitem{Scott94}
D.~R. Scott, J. Math. Phys. \text{35}, 5831 (1994), hep-th/9403030.

\end{thebibliography}
\end{document}